\documentclass{article}
\usepackage{spconf,amsmath,graphicx}
\usepackage{booktabs}
\usepackage{multirow}
\usepackage{hyperref}  
\usepackage{tablefootnote}
\usepackage{enumitem}

\title{TriNet: Stabilizing self-supervised learning from complete or slow collapse on ASR}
\name{Lixin Cao $^{1}$$^{\dagger}$  \qquad Jun Wang $^{1}$$^{\dagger}$ \thanks{${\dagger}$ Equal Contribution.} \qquad Ben Yang $^{1,2}$$^{\ddagger}$ \thanks{${\ddagger}$ Contribution made during internship in Tencent} \qquad Dan Su$^{1}$ \qquad Dong Yu$^{3}$}
\address{$^{1}$Tencent AI Lab, China \qquad $^{2}$ Peking University\qquad$^{3}$Tencent AI Lab, USA}

\begin{document}
%
\maketitle
\begin{abstract}
Self-supervised learning (SSL) models confront challenges of abrupt informational collapse or slow dimensional collapse. We propose TriNet, which introduces a novel triple-branch architecture for preventing collapse and stabilizing the pre-training. TriNet learns the SSL latent embedding space
and incorporates it to a higher level space for predicting pseudo target vectors generated by a frozen teacher. Our experimental results show that the proposed method notably stabilizes 
and accelerates pre-training and achieves a relative word error rate reduction (WERR) of $6.06\%$ compared to the state-of-the-art (SOTA) Data2vec for a downstream benchmark ASR task. We will release our code at \url{https://github.com/tencent-ailab/}.
\end{abstract}
\section{Introduction}
\label{sec:intro}

Self-supervised learning (SSL) models leverage unlabeled data, which makes significant advances \cite{chen_simclr2_20} and reaches performances almost on par with supervised baselines on many downstream tasks such as speech processing \cite{data2vec,wav2vec2,hubert21}. Among these models, state-of-the-art \textit{contrastive learning} methods \cite{wav2vec2,chen_simclr20,chen_simclr2_20,speech_simclr21} learn to reduce the distance between positive pairs of a sample and its distorted version, while increasing the distance between negative pairs of different samples.
They yield good performance with large amounts of contrastive pairs\cite{chen_simclr2_20}, which are difficult to mine and computationally intensive for training.

These challenges motivate alternative methods.
\textit{Bootstrapping} approaches \cite{BYOL_20,data2vec,simsiam_21} emerge to avoid using negative examples. Two networks are used to predict the same representation from augmented pairs. One is the teacher network with a stop-gradient (SG) operation (otherwise, a complete informational collapse may happen where the learned representations would rapidly collapse towards a single vector regardless of the inputs), and the other is the student network updating online.
Among these approaches, SimSiam \cite{simsiam_21} simply copied the student network's weights over to the teacher network; BYOL \cite{BYOL_20} updated the teacher network by tracking the exponential moving average (EMA) of the student network's weights. Data2vec\cite{data2vec} also took EMA to update the teacher network, but it used a masking prediction task similar to Wav2vec2\cite{wav2vec2} by feeding the student network with the masked data and the teacher network with the original data. Its objective is to predict the averaged embedding of several top layers of the teacher network, which is different from using only the top layer in BYOL.

As reported in Data2vec \cite{data2vec}, a collapse issue is more pronounced for speech tasks than computer vision or natural language processing tasks, due to the very correlated adjacent targets of the speech modality. It may come from two different natures \cite{collapse_22}: 1) the complete collapse; 2) a slow collapse like the observation made in \cite{vicreg_22} that the architectural tricks such as BYOL, Data2vec, and SiaSiam are not perfectly maintaining the variance of the representations, i.e., very slow collapse is still happening with these methods. Given these challenges, we are motivated to study novel regularization methods that are effective and practical for SSL models that are susceptible to complete or slow collapse. Hence, we propose a novel network TriNet, an analogy with a three-legged stabilizing stand ``Trivet", with following  contributions:
\begin{itemize}[leftmargin=*]
\item In contrast to most other pseudo-labeling approaches, TriNet does not require techniques such as K-means clustering, frame-level alignment, etc. For example, unlike Hubert \cite{hubert21}, which builds a fixed set of discrete target units by clustering, TriNet learns the SSL latent embedding space and incorporates it to a higher level space for predicting pseudo target vectors generated by a frozen teacher.
\item Not requiring to distract from any negative samples like Wav2vec2, Wav2vec-C\cite{wav2vec-c_21} or Spiral\cite{spiral_22} do, nor requiring any statistical assumption as the other advanced regularization approaches do (such as decorrelation \cite{vicreg_22} or maximizing log determinant \cite{CorInfoMax_22} which may not always be tenable for the sequences and tasks at hand), TriNet instead employs a third branch to generate stable and stale target vectors from the sequences themselves in the high-level space to construct regularization loss, which acts effectively as barriers against embedding space degeneracy.
\item Our experiment show that the proposed method stabilizes and accelerates \footnote{Comparing the pretraining time of SSL model with the frozen teacher to that without, while not counting the training of the frozen teacher model.} the pre-training and leads to significant performance improvements, with no requirement for more data augmentation or larger model capacities. 
\end{itemize}
Meanwhile, we would like to point out that TriNet achieves the above advances provided a frozen teacher model, although TriNet will notably surpass the frozen teacher, as we will demonstrate in the experiment. 
\section{Related Work}
\label{sec:prior}
Aside from contrastive methods for preventing informational collapse, 
other main trends are regularization methods for maximizing the information content of the embedding to prevent collapse. Recently, various regularization approaches are proposed to prevent the collapse in which the embedding variables contain highly redundant information. Among them, W-MSE\cite{WMSE_21}, Barlow-Twinss\cite{BarlowTwins_21}, and VICReg\cite{vicreg_22} attempt to produce embedding variables that are decorrelated from each other, whereas 
CorInfoMax\cite{CorInfoMax_22} does not constrain the variables to be uncorrelated but instead avoids covariance matrix degeneracy by using log-determinant as a regularizer loss function.
However, recent investigations show that these regularization terms worked effectively only if given specific SSL structural settings \cite{vicreg_22} and strong data augmentation \cite{lepage_22}. 
Note that all these regularization methods \cite{vicreg_22, CorInfoMax_22, WMSE_21, BarlowTwins_21} adopt an SSL-no-SG structure, where ``no-SG" means the  branch networks are both learnable with no stop-gradient. Instead, optimization of some regularization terms together with SSL-SG structures (\cite{BYOL_20,simsiam_21}) was found hard\cite{vicreg_22}. We also empirically observed that adding covariance regularization terms was not as effective in an SSL-SG structure. 
Data2vec\cite{data2vec} employs the SSL-SG structural tricks akin to BYOL\cite{BYOL_20} and Simsiam \cite{simsiam_21} that rely on a mechanism of normalizing the target to prevent collapse. This strategy seems effective but difficult to interpret and may lead to instabilities during the training\cite{vicreg_22, data2vec}. 

Our idea is also related to a different research area on pseudo-labeling.
BEST-RQ \cite{BEST-RQ_22} employs a random-projection quantizer to generate discrete pseudo labels. Hubert\cite{hubert21} uses an offline K-means clustering step to provide discrete pseudo labels for the masked regions,
and takes an iterative re-clustering and re-training process. These pseudo-labeling methods simplify the SSL targets to the level of clusters but essentially require the downstream tasks to be at the appropriate clustering level for the model to learn well. Another related idea is a combination of SSL and self-training \cite{self_train_1,self_train_2,self_train_3,self_train_4}. A fine-tuned SSL model \cite{SSL_self_1, SSL_self_2} or a supervised teacher model \cite{cheng_22} is used as the initial teacher model for pseudo-labeling the unlabeled set. Then a student model is trained on the combined labeled and pseudo-labeled data. 

Prior works on SSL stabilization including Wav2vec-C\cite{wav2vec-c_21} and Spiral\cite{spiral_22} are based on contrastive mechanisms, which prevent collapse by maximizing the distance between negative pairs. Non-contrastive approaches, which TriNet addresses, have collapsed minima independent of the input. Hence the fundamental question arises: how do multiple factors, like stop-gradients, EMA, teacher networks, and regularization, all come into play to avoid collapse? This leads to experimental studies like our TriNet and theoretical studies like \cite{yuandong_21,pokle_22}. Moreover, Wav2vec-C ``maintains a consistency towards the input features, of which the motivation is to facilitate codebook learning...by reconstructing the discrete codes to the input features.'' In contrast, TriNet predicts towards a high-level target space, of which the motivation is to stabilize the learning of the contextualized latent representation. Hence we don't require codebook learning, quantization, or VQ-specific loss algorithm for the codebook. SPIRAL proposed in-utterance contrastive loss\& position randomization to avoid model collapse\& positional collapse that arise on its own specific SG teacher design but not on the nonlinear learning dynamics of non-contrastive SSL\cite{yuandong_21}.

\begin{figure}[tb]
\centering
\includegraphics[width=\linewidth, scale=1.0]{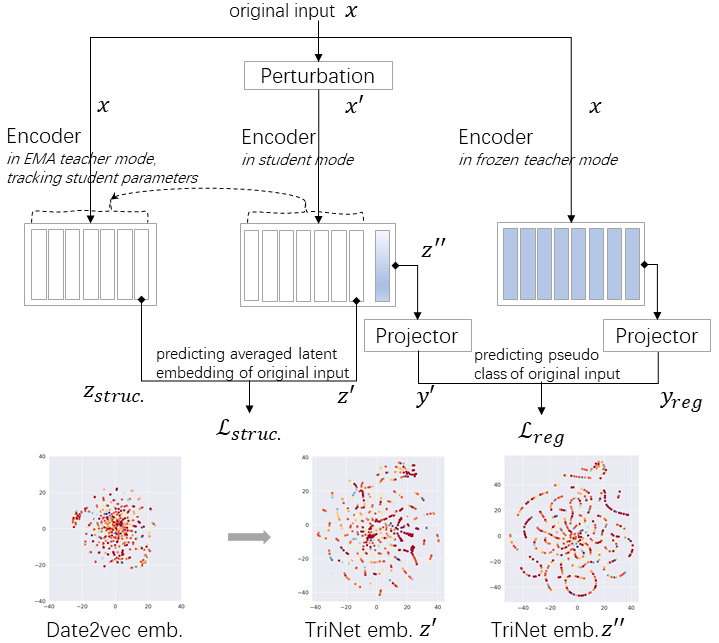}
\caption{TriNet with a three-legged structure: the left and right teacher networks perform in different modes to produce representations based on the original input, which are then predicted by the same middle network in student mode based on a perturbed version of the input. Bottom is t-SNE visualization of latent embedding of Data2vec and TriNet.}
\label{fig:network}
\vspace{-0.3cm}
\end{figure}
\section{Method}
\label{sec:method}
\subsection{Network Architecture}
\label{ssec:Network}
As illustrated on top of Fig. \ref{fig:network}, the proposed TriNet consists of three supporting networks. The middle ``leg" represents a student network that simultaneously regresses and predicts targets from the left and right teacher networks. The left teacher tracks the student parameters and generates the regression target, while the right teacher is a frozen fine-tuned model for automatic speech recognition (ASR) to generate high-level target vectors for stabilizing the whole training. Due to the different nature of the targets, we project both the student and the right teacher's embedding to a pseudo-class space.

We mask spans of the input sequence $x$ to generate the perturbed sequence $x'$ and feed it to a standard Conformer encoder \cite{fairseq} of the student.
The target $z_{struc.}$ is constructed by encoding the intact input $x$ with the same network but parameterized as an EMA teacher, as shown as the left ``leg" in Fig. \ref{fig:network}, and summarizing the teacher's top-K layer outputs \cite{caron21,BYOL_20}. This ``leg" adopts the same SSL structure as in Data2vec\cite{data2vec} and BYOL\cite{BYOL_20} for a straightforward comparison in this paper, whereas alternative SSL structures should be equally applicable. Meanwhile, TriNet stabilizes the training by introducing the third ``leg", as shown as the right branch in Fig. \ref{fig:network}, which takes the fine-tuned teacher to encode the intact input $x$ and generates pseudo target $y_{regul.}$ of the original input data. 
The design prevents the rest joint embedding architectures from abrupt or very slow collapse, in which output vectors produced by the branches are identical and constant, or end up spanning a low-dimension subspace.
\subsection{Pre-training}
\label{ssec:training}
In the proposed TriNet, we pretrain the student encoder to simultaneously learn contextualized representations of different levels and structural natures. The EMA teacher relies on the structural tricks of averaging (including both the moving average of model weights and the averaging of top-K layer outputs) to keep the prediction targets relatively stable while allowing the student to evolve freely and hopefully learn mid-level contextual representations. This freedom is a double-edged sword though-- the downside is that once the student starts to collapse, the EMA teacher will end up collapsing albeit very slowly (as demonstrated in Fig \ref{fig:losses}). 

To address either abrupt or slow collapse, 
the third branch plays the important role of ``anchor" by regularizing and avoiding cases in which the student and the EMA teacher degenerate together. TriNet employs the frozen teacher to provide high-level targets for regularization in a pseudo-class space, which is different from the mid-level embedding space between the student and the EMA teacher that maintains the SSL property. At the bottom of Fig. \ref{fig:network}, we use t-SNE \cite{tsne_08} to visualize the latent embedding generated by Data2vec and TriNet. Each point denotes a sample in a random batch and its color denotes its class. It indicates that TriNet actually arranges intra-class samples, of which the layout gets more obvious from $z'$ of the mid-level embedding space to $z''$ for the higher-level space, while the inter-class samples scatter all over the spaces.
\subsection{Regularization}
\label{ssec:objective}
Given the predicted latent embedding $z'$ in the mid-level embedding space and the contextualized targets $z_{struc.}$, we use a squared L2 norm loss to regress these targets:
\begin{align}
\mathcal{L}_{struc.}=\frac{1}{\sqrt{D}}\sum_{B\times T\times D}(z'_{n} - (z_{struc.})_{n})^2,
\end{align}
where $n$ is the index of a total of $B\times T\times D$ elements in a batch, and $B$, $T$, $D$ are batch, frame, and dimension sizes.

Given the prediction $y'$ in the high-level space and the pseudo-class targets $y_{regul.}$, we examine and compare two kinds of objectives. One is a squared L2 norm for regression:
\begin{align}
\mathcal{L}_{regre.}= \frac{1}{\sqrt{D}}\sum_{B\times T\times D}(y'_{n} - (y_{regul.})_{n})^2,
\end{align}
The other is a cross-entropy loss for classification:
\begin{align}
\mathcal{L}_{regul.}= \frac{1}{\sqrt{D}}\text{CrossEntropy}(y', \text{Softmax}(y_{regul.})),
\end{align}

Our ablation study shows that ${L}_{regul.}$ is more effective than ${L}_{regre.}$. We consider that is because, in the high-level space, ${L}_{regul.}$ is a more suitable measure of how well the predictions are made by a pseudo-phoneme classification rather than a latent embedding regression, again echoing the difference of the various-level spaces and their complementary regularization effects. Consequently, we adopt $\mathcal{L}= \mathcal{L}_{struc.}+ \mathcal{L}_{regul.}$ as training objective in our experiments.

\section{Experiments}
\label{sec:experiments}
\subsection{Pre-training and Fine-tuning}
\label{ssec:exp-pre-training}
We pre-train models on Librispeech \cite{pana15} that contains 960 hours of speech (LS-960h), and fine-tune for ASR on the clean 100h (LS-100h) subset of LS-960h. 
We also pre-train on a much larger dataset Libri-light (LL-21k)\footnote{We took the samples less than $32$s and constructed a 20935-hour subset.} \cite{librilight_20} and fine-tune for ASR on LS-960h. We evaluate the standard Librispeech dev-clean/other and test-clean/other sets.

We implemented the reference Data2vec Base and Large models based on Fairseq\cite{fairseq} and the proposed model TriNet with the same corresponding backbone architectures, dropout and masking strategies (see detailed configurations in \cite{wav2vec2}). To save memory footprint and for fair comparison, all models apply the same pre-processing: the input 16 kHz waveform is first transformed into an 80-dim filter bank than the raw waveform; it is then processed by a feature extractor containing two Convolution-2D subsampling layers with $576$ channels, strides (2,2), and kernel widths (3,3). This results in an output sequence of $1/4$ of the original length. The input is applied with layer norm before sending to the encoder.
Other hyper-parameters, including annealing rates, optimizers, learning rate schedulers, and fine-tuning regimes, also follow \cite{data2vec} and otherwise would be described if different.

In our experiments, we used the frozen teacher fine-tuned on LS-100h and LS-960h for the LS-960h and LL-21k unlabelled data pretraining, respectively, to demonstrate that TriNet with no requirement for additional data, larger model capacity, or varying model structural tricks will surpass the frozen teacher. For pre-training, TriNet uses all but the last Conformer blocks for encoding the mid-level latent embedding space illustrated as blank blocks in Fig. \ref{fig:network}, and dedicates the last Conformer block to the high-level space (blue blocks in Fig. \ref{fig:network}). The overall learnable model sizes are identical to Data2vec Base and Large. While an arbitrary teacher with a heterogeneous architecture or a single pass using pre-generated pseudo targets is also applicable for TriNet, we leave that part of investigation for future work. 

\subsection{Results}
\label{sec:results}
\begin{figure}[tb]
\centering
\setlength{\abovecaptionskip}{0cm}
\includegraphics[width=0.85\linewidth, scale=1.0]{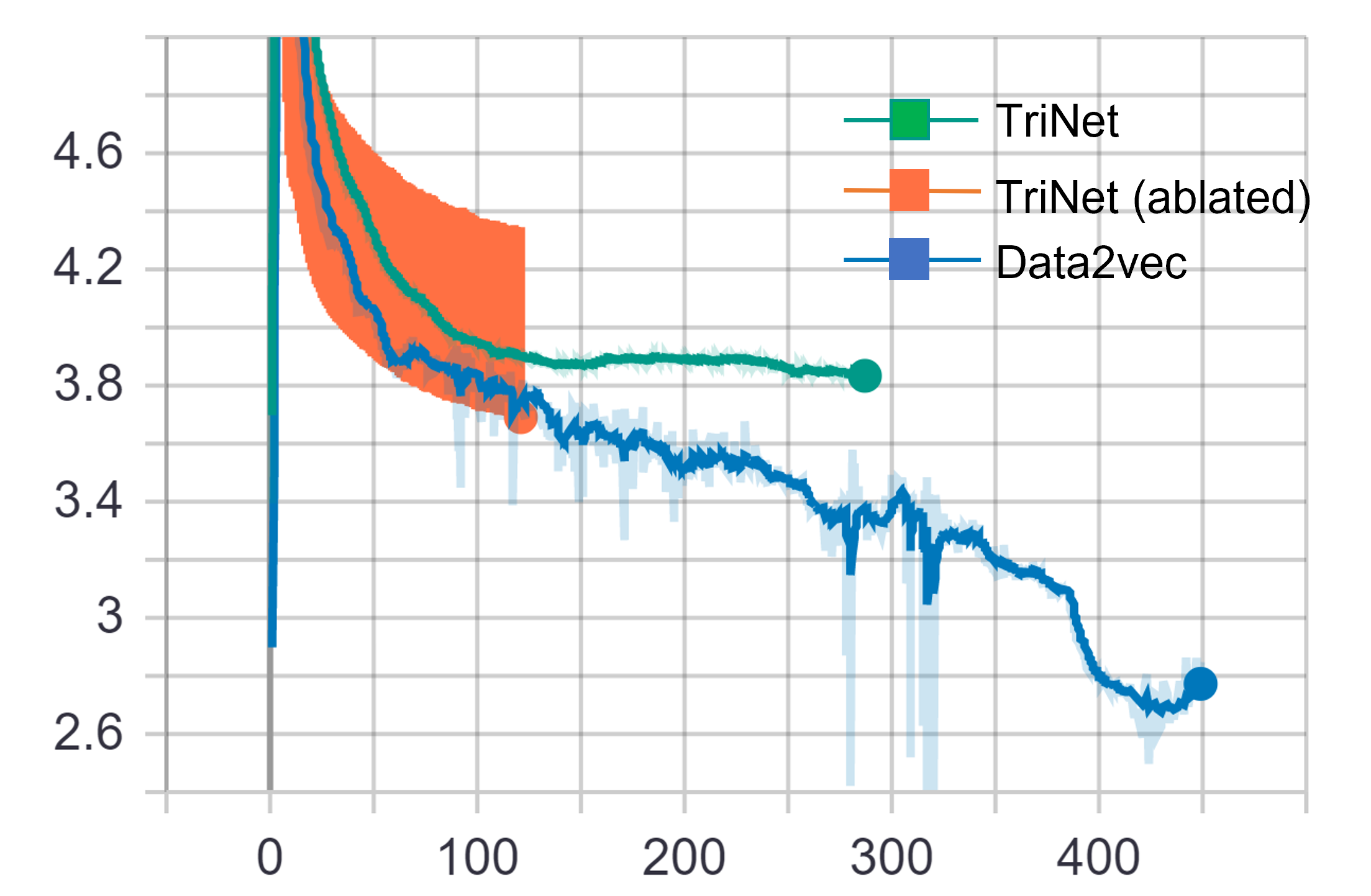}
\caption{Pre-training losses}
\label{fig:losses}
\vspace{-0.5cm}
\end{figure}

In Fig. \ref{fig:losses}, the drop of the Data2vec loss from epoch 350 to 450 actually reflected a slow collapsing case --- the downstream fine-tuned model has the word error rate (WER, via greedy search on the dev-other set) degenerating from $9.949\%$ at epoch 300 to $10.25\%$ at epoch 350 and further to $10.432\%$ at epoch 400. We chose the best Data2vec checkpoint at epoch 300 for the following comparison.

Meanwhile, we can observe the pre-training loss of TriNet is much smoother and stabler than that of Data2vec (light blue curve is without smoothing. Note the absolute loss values are not directly comparable), indicating that the frozen teacher in TriNet works effectively as an ``anchor" by providing stable and stale regularization and preventing the EMA teacher and the student from drifting together toward a collapsed subspace. Meanwhile, TriNet manages to converge within  $2/3$ of the overall epochs of Data2vec.


Pre-trained models are fine-tuned on LS-100h labeled for ASR by mapping the representations via a randomly initialized linear projection on top of the network into $32$ classes representing the vocabulary. Models are optimized by minimizing a CTC\cite{ctc_06} loss. 

As shown in Table \ref{tab:wer}, our approach achieves relative word error rate reductions (WERRs) of $9.58\%/3.53\%/6.47\%/4.67\%$ ($6.06\%$ in average) over Data2vec on dev-clean/other and test-clean/other of the Librispeech benchmark. Moreover, it achieves WERRs of $6.01\%/7.33\%/4.5\%/5.67\%$ ($5.89\%$ in average) when pretraining on the much larger unlabeled LL-21k, reflecting the scalability of the proposed method.
\begin{table}[t]
\centering
\caption{WER ($\%$) on the Librispeech dev/test sets when pretrained on LS-960h \& LL-21k unlabeled and finetuned on the LS-100h \& LS-960h labeled, respectively.}
\label{tab:wer}
\begin{tabular}{lccccc}
 \multirow{2}{*}{Model} & \multirow{2}{*}{\shortstack{Unlabled \\ data}} & \multicolumn{2}{c}{dev} & \multicolumn{2}{c}{test} \\
\cmidrule(lr){3-4}\cmidrule(lr){5-6}
&  & clean & other & clean & other \\
 \hline\hline
 Wav2vec2\cite{wav2vec2} & LS-960h & 2.7 & 7.9 & 3.4 & 8.0 \\
 Hubert\cite{hubert21}& LS-960h  & 2.6 & 7.8 & 3.4 & 8.1 \\
 Data2vec & LS-960h & 2.61 & 6.79 & 3.09 & 7.07 \\
 \midrule
 TriNet (ablated) & LS-960h & 2.49 & 7.16 & 2.95 & 7.23 \\
 TriNet  & LS-960h & {\bf{2.36}}& {\bf{6.55}} & {\bf{2.89}} & {\bf{6.74}} \\
 \hline\hline
 Data2vec & LL-21k & 1.83 & 4.91 & 2.42 & 5.29 \\
  TriNet  & LL-21k & {\bf{1.72}}& {\bf{4.55}} & {\bf{2.31}} & {\bf{4.99}} \\
 \bottomrule
\end{tabular}
\vspace{-0.5cm}
\end{table}

\subsection{Ablations}
\label{ssec:ablation}
To exam the different natures of the two spaces, we make an ablation study by removing the projectors and spare no Conformer layer specific for the high-level target (blue blocks in Fig. \ref{fig:network}). It turns out the training becomes rather unstable, as shown as the orange curve of the training loss in Fig. \ref{fig:losses}, indicating the learning process is dragged zig-zag between the two spaces of different natures and can not converge well.

Another ablation study is on comparing the regularization terms of ${L}_{regul.}$ and ${L}_{regre.}$. The second line from the bottom of Table \ref{tab:wer} indicates the result by replacing ${L}_{regul.}$ with ${L}_{regre.}$. Although the result marginally outperforms Data2vec, it is much worse than TriNet before ablation. This validates the effectiveness of ${L}_{regul.}$ being a more suitable measure for the high-level space than an MSE loss that is suitable for mid-level embedding regression, reflecting the complementary nature of the spaces constructed at different levels via the triple ``legs" in TriNet.
\section{Conclusion}
\label{sec:conclusion}
The proposed TriNet addresses challenges of complete or slow collapse for SSL architectures and shows efficacy on the downstream ASR tasks.
TriNet employs a novel architecture that utilizes a frozen teacher to generate pseudo targets as ``anchors" for stabilizing the remaining part of the joint embedding SSL architecture. Besides, it succeeds in accelerating the pre-training and obtaining significant WERR compared to the SOTA Data2vec model in benchmark ASR tasks.
\vfill\pagebreak

\small
\bibliographystyle{IEEEbib}
\bibliography{draft}

\end{document}